\documentclass[aps,prl,reprint,groupedaddress]{revtex4-1}
\usepackage{graphicx}
\usepackage{amsmath}
\usepackage{xcolor}

\draft 

\def\rc{$\alpha$-RuCl$_{3}$}

\begin{document}


\title{\textit{J}$_{eff}$ description of the honeycomb Mott insulator \rc}

\author{A. Koitzsch,$^{1}$ C. Habenicht,$^{1}$ E. M\"uller,$^{1}$ M. Knupfer,$^{1}$ B. B\"uchner,$^{1}$ H. Kandpal,$^{1}$ J. van den Brink,$^{1}$ D. Nowak,$^{2}$ A. Isaeva,$^{2}$ and Th. Doert$^{2}$}

\affiliation
{$^{1}$IFW-Dresden, P.O.Box 270116, D-01171 Dresden, Germany\\
$^{2}$Technische Universit\"at Dresden, Department of Chemistry and Food Chemistry, D - 01062 Dresden, Germany\\}
\date{\today}

\begin{abstract}
Novel ground states might be realized in honeycomb lattices with strong spin--orbit coupling. Here we study the electronic structure of \rc, in which the Ru ions are in a \textit{d}$^5$ configuration and form a honeycomb lattice, by angle-resolved photoemission, x-ray photoemission and electron energy loss spectroscopy supported by density functional theory and multiplet calculations. We find that \rc\ is a Mott insulator with significant spin-orbit coupling, whose low energy electronic structure is naturally mapped onto \textit{J}$_{eff}$ states. This makes \rc\ a promising candidate for the realization of Kitaev physics. Relevant electronic parameters such as the Hubbard energy \textit{U}, the crystal field splitting 10\textit{Dq} and the charge transfer energy $\Delta$ are evaluated.
Furthermore, we observe significant Cl photodesorption with time, which must be taken into account when interpreting photoemission and other surface sensitive experiments. 
\end{abstract}


\maketitle 

The search for novel electronic and magnetic groundstates has ever been a driving force of condensed matter physics. The effects of strong spin-orbit coupling, possibly competing with other energy scales, have turned out to be especially fruitful in this respect in recent years. This is most prominently manifested by the advent of topological insulators \cite{Hasan2010}. More recently, the Kitaev model was established, which describes the bond-dependent spin interactions on a honeycomb spin 1/2 lattice \cite{Kitaev2006}. The Kitaev model attracts enormous attention because it is exactly solvable and its ground state is an exotic quantum spin liquid. However, unambiguous experimental evidence is lacking so far. The prime candidates for the realization of Kitaev physics have been the 5\textit{d}$^5$ iridates \textit{A}$_2$IrO$_3$ (\textit{A} = Na, Li) \cite{Chaloupka2010, Kim2008, Singh2010, Liu2011, Choi2012}. This thread of research relies on the realization of effective J$_{eff}$ = 1/2 pseudospins by the combined interaction of spin-orbit coupling and crystal field splitting.   
But the concept of J$_{eff}$ = 1/2 pseudospins is under debate for the iridates due to substantial lattice distortions lifting the \textit{t}$_{2g}$ degeneracy, which, strictly speaking, invalidates the J$_{eff}$ description.

\rc\ appeared recently against this background as a 4\textit{d} analogue to the iridates \cite{Plumb2014, Banerjee2015}. Ru is in a 3+ state and features a \textit{d}$^5$ electron count with a low spin state. Its spin-orbit coupling ($\lambda$ $\approx$ 0.1 eV) is strongly reduced as compared to the iridates, but so is its bandwidth \textit{W} due to presumed correlation effects. Importantly, the local cubic symmetry is almost perfect in contrast to the iridates. Hence, the J$_{eff}$ description might be still operable for \rc. Another practical advantage is that it can be synthesized as large, easy-to-cleave single crystals, which offer the possibility of exfoliation.

RuCl$_3$ has been known for a long time and is even of some importance as a chemical \cite{Bose1928}.
Its electronic structure has been repeatedly investigated over the years by optical spectroscopy and photoemission \cite{Binotto1971, Pollini1994, Pollini1996}. 
The picture of a Mott-insulating state was proposed where the Ru 4\textit{d} bands are situated close to \textit{E}$_F$ but show little dispersion \cite{Pollini1996}. More recent optical data confirmed the magnitude of the charge gap \textit{E}$_G$ $\approx$ 1.1 eV, which is much smaller than the charge transfer energy $\Delta$ $\approx$ 5 eV as expected for a Mott insulator \cite{Sandilands2015}. \textit{U} has been estimated to be about 1.5 eV \cite{Pollini1996, Plumb2014}, a value often used as an input parameter for bandstructure calculations. 

As for the magnetic properties, one or two (depending on the study) phase transitions at $T\approx7$ K and 15 K are reported \cite{Majumder2015, Sears2015, Johnson2015}. A strong magnetic anisotropy is found with large effective moments exceeding the S = 1/2 limit and implying a large orbital contribution. Neutron scattering below \textit{T} = 7 K is consistent with a zigzag type order, one of the types of magnetic order predicted within the framework of the Kitaev - Heisenberg model \cite{Banerjee2015, Johnson2015}.
 
Having mentioned some of the favorable properties of \rc\ above, a disadvantage of the material might be its sensitivity to environment and treatment, which may contribute to a certain scatter of reported physical properties. The crystal structure is susceptible to stacking faults  \cite{Johnson2015} and other defects. 
We will show that another point must be added here, namely substantial Cl photodesorption at the surface.  

Nevertheless, carefully taking into account this complication, we could elucidate the electronic structure of \rc\ by state of the art photoemission (PES), electron-energy-loss spectroscopy (EELS), density-functional-theory (DFT) and multiplet calculations.
We achieve a consistent, quantitative picture of a spin-orbit assisted Mott-insulator. The central question of this study, and decisive for the prospects of \rc\ as a possible carrier of Kitaev groundstates, is whether or not the J$_{eff}$ = 1/2 description of the electronic structure is appropriate. Based on the comparison of the DFT calculations with results from angle-resolved photoemission spectroscopy (ARPES), we can answer this question affirmatively.

Platelet-like single crystals up to several mm in diameter of \rc\ were grown by chemical vapor transport reactions.
PES measurements were performed using a laboratory based system at room temperature after cleaving the crystals \textit{in situ}.
The EELS (Electron Energy Loss Spectroscopy) measurements in transmission have been conducted using thin films ($\textit{d} \approx$ 100 -- 200 nm) at $T=20$ K.
The density functional calculations were performed using the all-electron full-potential local-orbital (FPLO) code \cite{Koepernik1999, FPLO} within Perdew-Wang parametrization \cite{Perdew1992}. 

See the supplemental material for details.


\begin{figure}[t]
\includegraphics[width=1\linewidth]{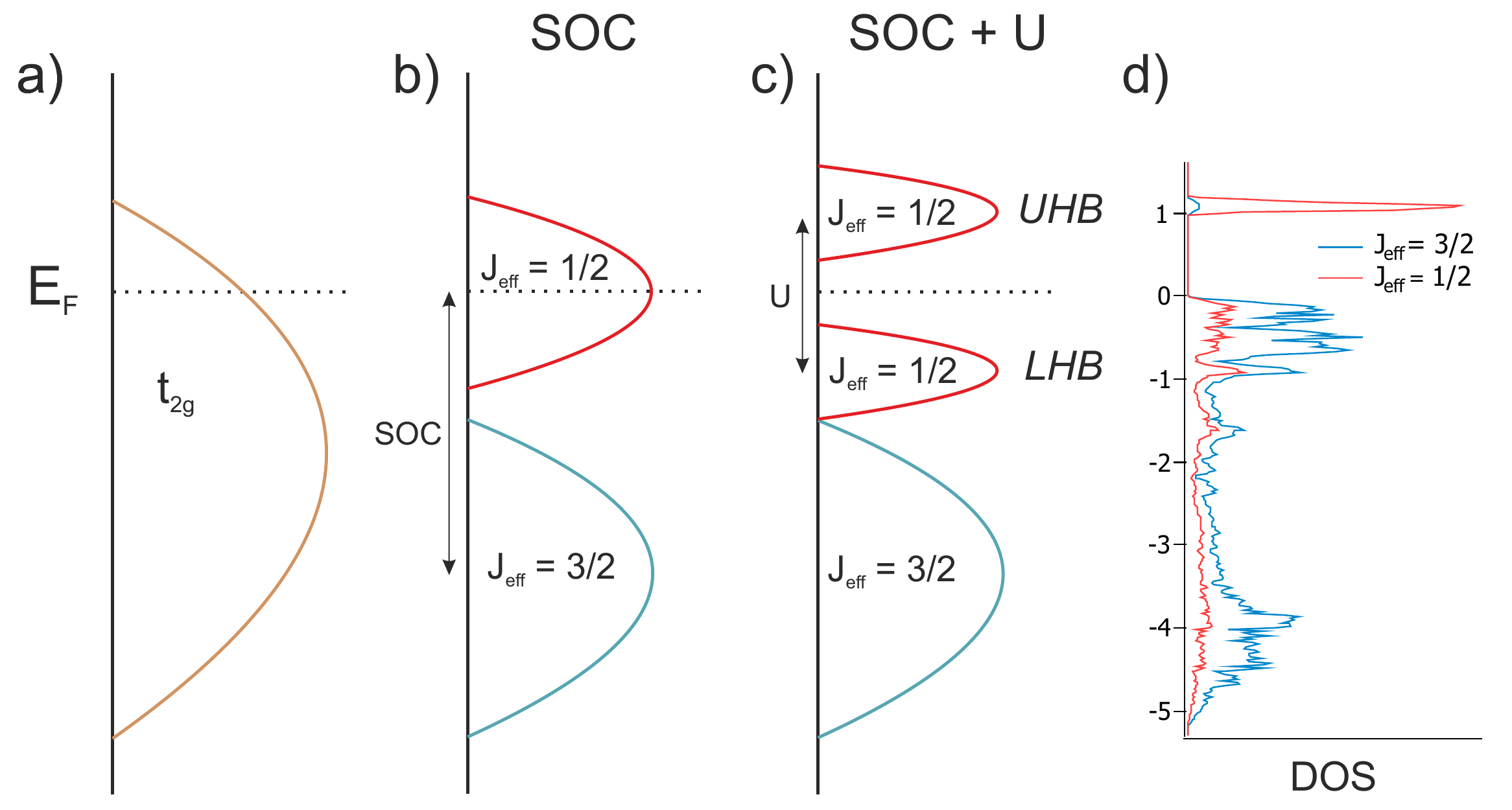}%
\caption{(a - c) \textit{J$_{eff}$} description of the \textit{d} - level electronic structure. a) Schematic density of states without interactions. b) Under the presence of strong spin-orbit coupling. c) With spin-orbit coupling and on-site correlation \textit{U}. d) Calculated density of states of \rc\ with spin-orbit coupling and on-site correlation.\label{}}%
\end{figure}

Under a cubic crystal field, the 4\textit{d} electron manifold of Ru splits into \textit{t$_{2g}$} and \textit{e$_{g}$} states separated by the crystal field splitting parameter \textit{10Dq}. Fig. 1a shows schematically the \textit{t$_{2g}$} states. This band cuts the Fermi energy (\textit{E$_F$}), because in a low-spin configuration with 5 \textit{d}-electrons it is not completely filled. 
Introducing the spin-orbit coupling, the $J_{eff} = 1/2$ and 3/2 states separate from each other (Fig. 1b). The on-site correlation energy \textit{U} causes a gap opening within the $J_{eff} = 1/2$ band. Fig. 1d shows the density of states (DOS) of \rc\ obtained by DFT calculated by taking into account SOC and the correlation energy ($U=2$ eV, J$_{H}=0.4$ eV, \textit{U$_{eff}$} = 1.6 eV) projected onto \textit{J$_{eff}$} = 1/2 and 3/2. The DOS correctly reproduces the gap opening and the insulating nature of \rc. Comparing it to the general \textit{J$_{eff}$} picture, it bears out the almost pure $J_{eff} = 1/2$ character of the sharp upper Hubbard band (UHB). The lower Hubbard band (LHB), on the other hand, is strongly mixed with the $J_{eff} = 3/2$ states. This is a consequence of the antiferromagnetic order imposed on this calculation, which gives the lowest total energy.
The \textit{J$_{eff}$} description appears to be well justified by the DFT. In the following we compare the DFT to a variety of experimental probes. 

\begin{figure}[t]
\includegraphics[width=1\linewidth]{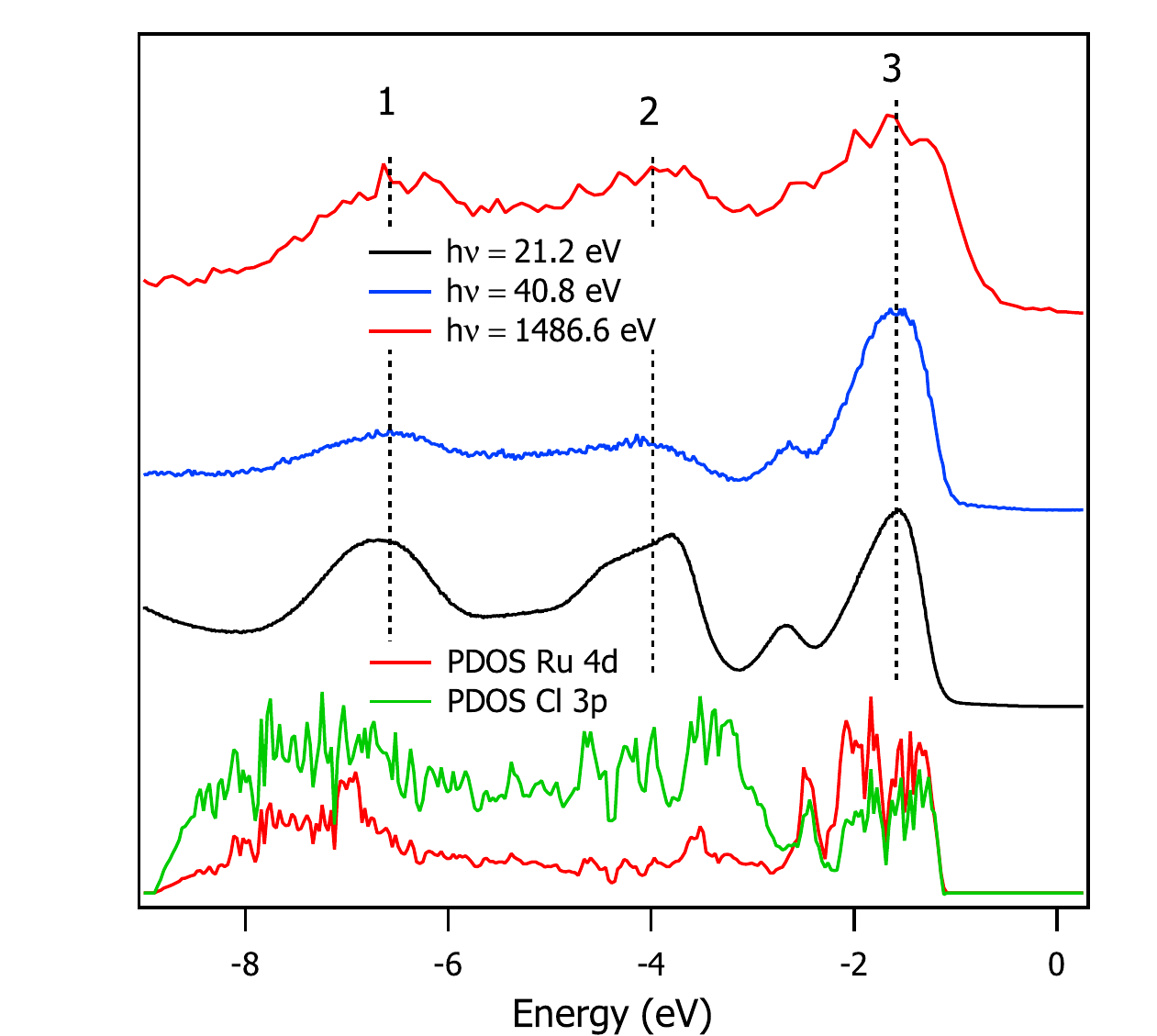}%
\caption{Valence band of \rc\ measured by photoemission spectroscopy with different photon energies at room temperature compared to an orbital projected density of states calculation.\label{}}%
\end{figure}

Fig. 2 shows angle-integrated photoemission spectra of the valence band region taken with three different photon energies. The spectra consist of three main features, labeled 1 - 3. Their relative intensity varies as a function of photon energy. Peaks 1 and 2 display an intensity minimum relative to peak 3 for \textit{h}$\nu$ = 40.8 eV. This reflects the different orbital character of the underlying states, which causes a different dependence of the photoionization cross section on photon energy. By comparison with tabulated values peak 1 and 2 can be assigned to Cl 3\textit{p} and peak 3 to Ru 4\textit{d} states \cite{sigma} in agreement with previous results \cite{Pollini1994}. However, an important difference to the earlier studies is that we observe a larger gap. The onset of the valence band is located at E$_{VBO}\approx$ 1 eV. Together with \textit{E}$_G$ $\approx$ 1.2 eV (see Fig. 3) this places the Fermi level close to the bottom of the \textit{d}$^6$ conduction band. The full width half maximum of the Ru 4\textit{d} peak (without the shoulder at \textit{E} = - 2.7 eV) is \textit{W$_{4d}$} = 0.75 eV only, which indicates that the system is susceptible to correlation effects even for moderate values of \textit{U}.

\begin{figure*}[t!]
\includegraphics[width=0.8\linewidth]{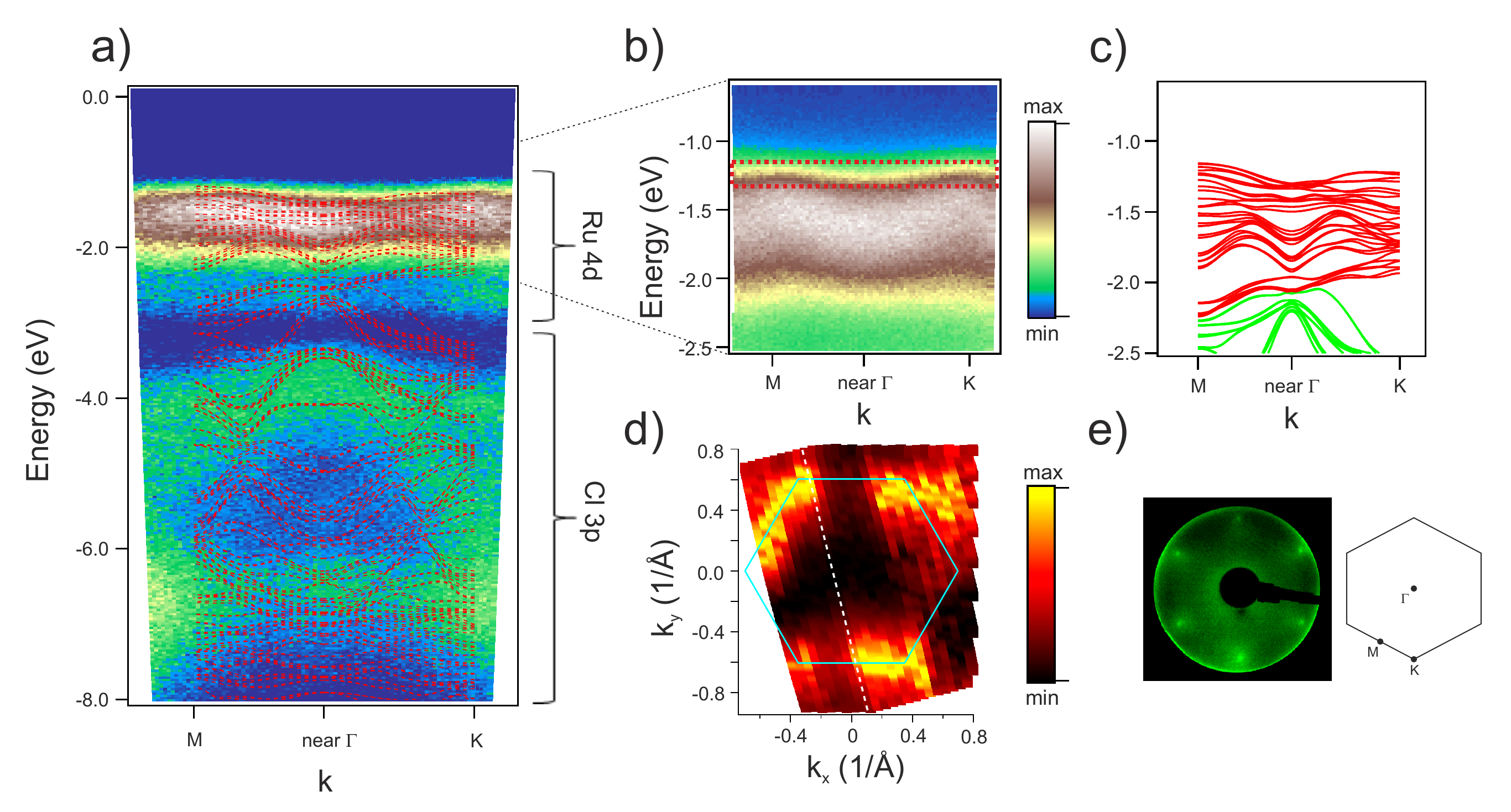}%
\caption{a) Color coded representation of the angle-dependence of the full valence band along the white line in panel (d). Energy regions of mainly Ru 4\textit{d} and Cl 3\textit{p} character are denoted on the right side. Red dotted lines are results of bandstructure calculations under the same renormalization as the DOS in Fig. 2. b) Expansion of the Ru 4\textit{d} region. The dashed red box indicates position and integration window of panel (d). c) DFT derived bandstructure without bandwidth renormalization but with offset. Red lines denote bands of mainly Ru 4\textit{d} character, green for Cl 3\textit{p}. d) Constant energy k$_x$ - k$_y$ map at \textit{E} = 1.2 eV. The integration window is highlighted in (b) by a red dashed box. e) Low-energy electron diffraction of a cleaved surface and surface Brillouin zone of \rc. }. 
\end{figure*}

The lower part of Fig. 2 presents the Ru 4\textit{d} and Cl 3\textit{p} orbital projected DOS. It is identical to the DOS in Fig. 1d but has been renormalized for better comparison with experiment (stretching factor 1.5, offset 1.1 eV). The low energy region is dominated by the Ru 4\textit{d} weight while at higher energies Cl 3\textit{p} dominates in agreement with experiment. Also the main features of the valence band (1 - 3) are well reproduced up to finer structures e.g. the \textit{E} = - 2.7 eV shoulder. 

Fig. 3a exhibits the angle-dependence of the valence band along a cut in \textbf{k} - space highlighted in Fig. 3d. The renormalized theoretical bandstructure is overlayed over the experimental data. Two regimes can be distinguished: the low-energy region of weakly dispersing Ru 4\textit{d} states and at higher energies more strongly dispersing Cl 3\textit{p} derived bands, which are clearly separated from each other. The calculation is again in qualitative agreement to the data. 
The low-energy region is expanded in Fig. 3b along the same cut. The Ru 4\textit{d} bands disperse on the order of 200 meV. They form a minimum around $\Gamma$ and maxima around \textit{M} and \textit{K}. Note, that this is not the dispersion of a single band but a superposition of many \textit{d} - bands forming a broad peak. In order to compare it to theory, we present the unrenormalized, but offset  low-energy bandstructure in Fig. 3c. The total bandwidth and the principal shape of the dispersion agree with experiment. 

The horizontal red box corresponds to the energy position and integration window of the k$_x$ - k$_y$ map presented in Fig. 3d. The contour of highest intensity follows the border of the surface Brillouin zone (SBZ) shown schematically in Fig. 3e. 
Fig. 3e shows a low-energy electron diffraction (LEED) image of a sample surface after finishing the photoemission measurements. It exhibits sharp spots with the expected symmetry. 

The photoemission experiments presented so far probe the occupied states only. The charge gap is not accessible in this way. We have, therefore, performed EELS experiments and measured the loss function, which can be expressed as Im(-1/$\epsilon$) where $\epsilon$ is the dielectric function (see Fig. 4).
The loss function shows several features in the low energy region. We assign peak A at \textit{E}$_G$ = 1.2 eV to interband transitions across the charge gap, i.e. \textit{d}$^5$\textit{d}$^5$ $\rightarrow$ \textit{d}$^4$\textit{d}$^6$. Its energy reflects \textit{U} reduced by half of the bandwidth of upper and lower Hubbard band. From \textit{E}$_G$ = 1.2 eV and \textit{W$_{4d}$} = 0.75 eV, \textit{U} = 1.6 eV is a reasonable estimate. Here, we have assumed the width of the UHB to be small. This is justified by the narrow line width of the A - feature and also by the DFT, which predicts a very narrow line (see Fig. 1d). In fact, the A feature is narrower than \textit{W$_{4d}$}, which implies that excitonic effects contribute to the gap excitation. 

Excitation B centered at \textit{E} = 2.1 eV can be associated with the cubic crystal field splitting, i.e. excitations into the \textit{e}$_{g}$ states and is approximately 10\textit{Dq}. This claim will be substantiated further when discussing the core levels (Fig. 5). Features C and D are both due to Cl 3\textit{p} - Ru 4\textit{d} charge transfer excitations. The onset of C is $\approx$ 3 eV which corresponds to the onset of peak 2 in Fig. 1. The energy scales of photoemission and energy loss agree roughly in this case because the \textit{d}$^6$ conduction band is situated close to \textit{E}$_F$. The energy difference of features C and D is equal to the separation of peaks 1 and 2 in Fig.1 lending further support to this assignment. Both features are broad and show fine structures reflecting the pronounced dispersion of the Cl 3\textit{p} bands. 

\begin{figure}[h]
\includegraphics[width=1\linewidth]{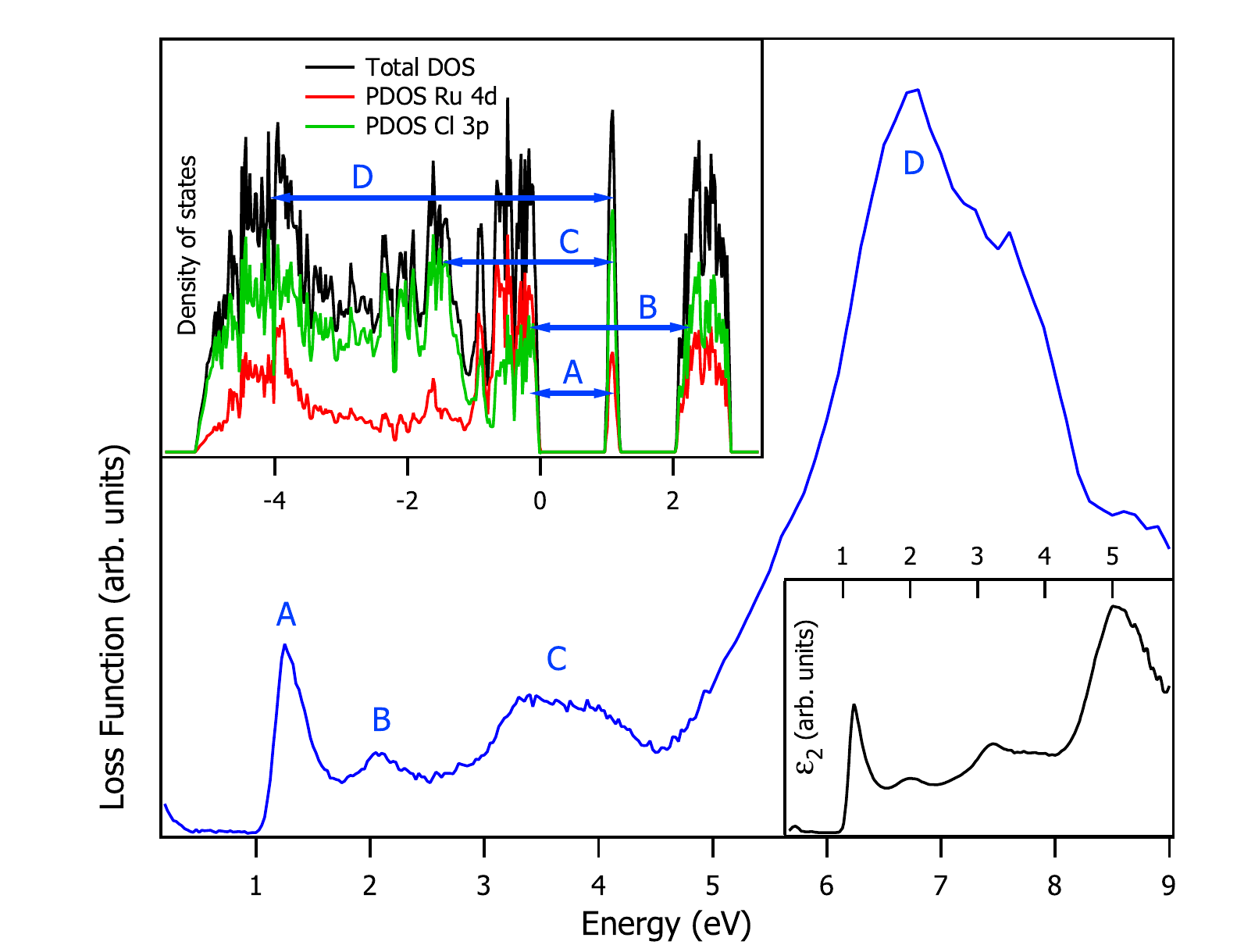}%
\caption{Loss function measured by EELS at \textbf{q} = 0.1 \AA. Lower inset: $\epsilon_2$ reproduced from Plumb et al. \cite{Plumb2014}. Upper inset: Orbital projected DOS with transitions labeled according to the peaks in the main panel. \label{}}%
\end{figure}

The lower inset of Fig. 4 reproduces $\epsilon_2$ extracted from optical measurements as published by Plumb et al. \cite{Plumb2014}. The agreement to the loss function is satisfying, although, they are not equivalent quantities, in particular when $\epsilon_1$ is large.
The upper inset shows again the DOS with an assignment of the transitions A -- D.

Taking together Figs. 2 - 4, a consistent description of the electronic structure of \rc\ is obtained. This is achieved by a theoretical framework, which itself can be naturally mapped onto the \textit{J}$_{eff}$ description. This implies that the latter is indeed meaningful for \rc.

The electronic parameters introduced so far (\textit{U}, $\Delta$, 10\textit{Dq}) determine the shape of the core levels, which can then be used to check the validity of the above arguments. 

We come now to the discussion of the core levels. Fig. 5 presents the 3\textit{p} spectra measured by XPS and EELS, along with multiplet calculations using a joined set of parameters. The spectra are split due to the 3\textit{p} spin-orbit interaction by 22 eV in a 3\textit{p}$_{3/2}$ and a 3\textit{p}$_{1/2}$ state. A double peak structure is observed in the EELS 3\textit{p}$_{3/2}$ line which is absent in 3\textit{p}$_{1/2}$ and in XPS. In fact, this double feature is a consequence of the spin-orbit coupling of the \textit{d} - electrons. The latter splits the \textit{d}-levels into 4\textit{d}$_{5/2}$ and 4\textit{d}$_{3/2}$ states. According to the \textit{J} selection rule, the 4\textit{d}$_{3/2}$ state can be reached by both 3\textit{p}$_{1/2}$ and 3\textit{p}$_{3/2}$, whereas the 4\textit{d}$_{5/2}$ is accessible for the 3\textit{p}$_{3/2}$ state only. The double peak of the 3\textit{p}$_{3/2}$ EELS line indicates that the \textit{t}$_{2g}$ hole is of \textit{J} = 5/2 character. The energy difference between the two components is mainly determined by the crystal field splitting.

For the photoemission, the transitions to \textit{t}$_{2g}$ or \textit{e}$_{g}$ are not operative and only one ionization peak is observed. Fig. 5 confirms that the spin-orbit coupling must be taken into account when describing the \textit{d}-electrons. A similar conclusion was drawn previously from the lineshape of the Ru L$_{2,3}$ edge of x-ray absorption spectra \cite{Plumb2014}.

\begin{figure}[h]
\includegraphics[width=1\linewidth]{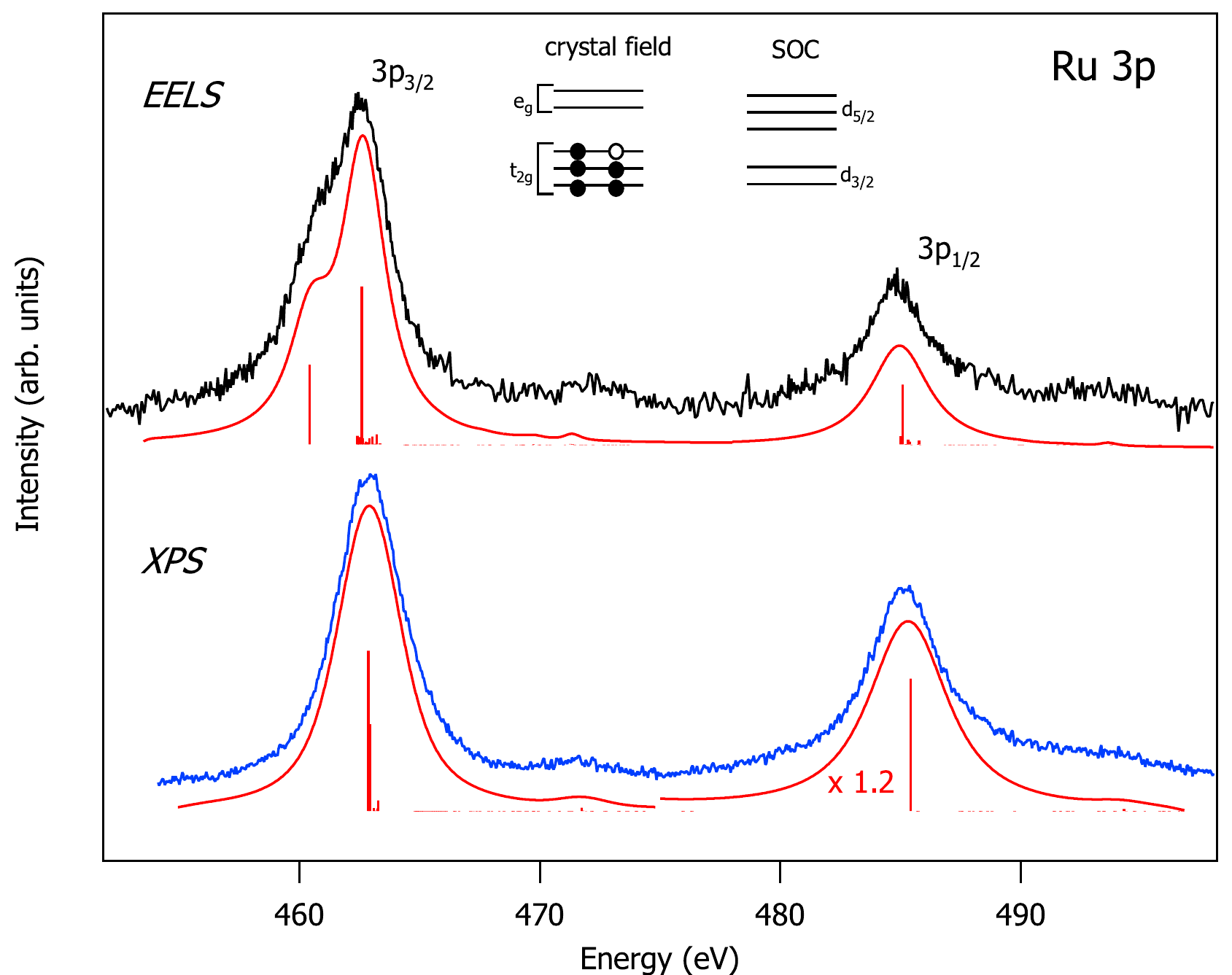}%
\caption{Ru 3p EELS and XPS. Red lines: charge-transfer multiplet calculations (see text for details).\label{}}%
\end{figure}

Even more information is hidden in Fig. 5. Closer inspection reveals weak satellite features at the high energy side of the main peaks with an energy separation of about 8 eV. Such structures are well known in transition metal compounds. They originate from charge transfer processes, where an electron from the surrounding Cl - ligands hops to the central Ru. In order to achieve a complete description and understanding of the data, we modeled the spectra by charge-transfer multiplet calculations using the \emph{CTM4XAS} package \cite{Stavitski2010}. Motivated by the low energy PES and EELS results discussed above, we fixed 10$Dq =2.2$ eV, $\Delta$ = 5 eV. The latter is to be considered as an average value for the broad energy region of charge transfer excitations seen in Fig. 4. The Slater integrals have been reduced to 25 \% of their atomic values \cite{deGroot1994}. We set the overlap integrals \textit{T}$_{e_g}$ = 2 eV and \textit{T}$_{t_{2g}}$ = 1 eV and the \textit{d}-level  core-hole repulsion \textit{U}$_{dc}$ = 3 eV. The latter is usually 1 - 2 eV larger than \textit{U}.
With these parameters and appropriate Gaussian and Lorentzian broadening the red spectra are obtained. For XPS, quantitative agreement is accomplished. The calculation reproduces the shape of the principal lines and position and weight of the satellites. The branching ratio differs slightly, probably due to cross section and diffraction effects.
The shape of the EELS spectrum is also correctly reproduced, in particular the line splitting of the \textit{p}$_{3/2}$ peak. The satellite intensity is underestimated by the calculation due to the presence of a Ru 3p $\rightarrow$ Ru 5s transition \cite{Sham1983}. 

\begin{figure}[h]
\includegraphics[width=1\linewidth]{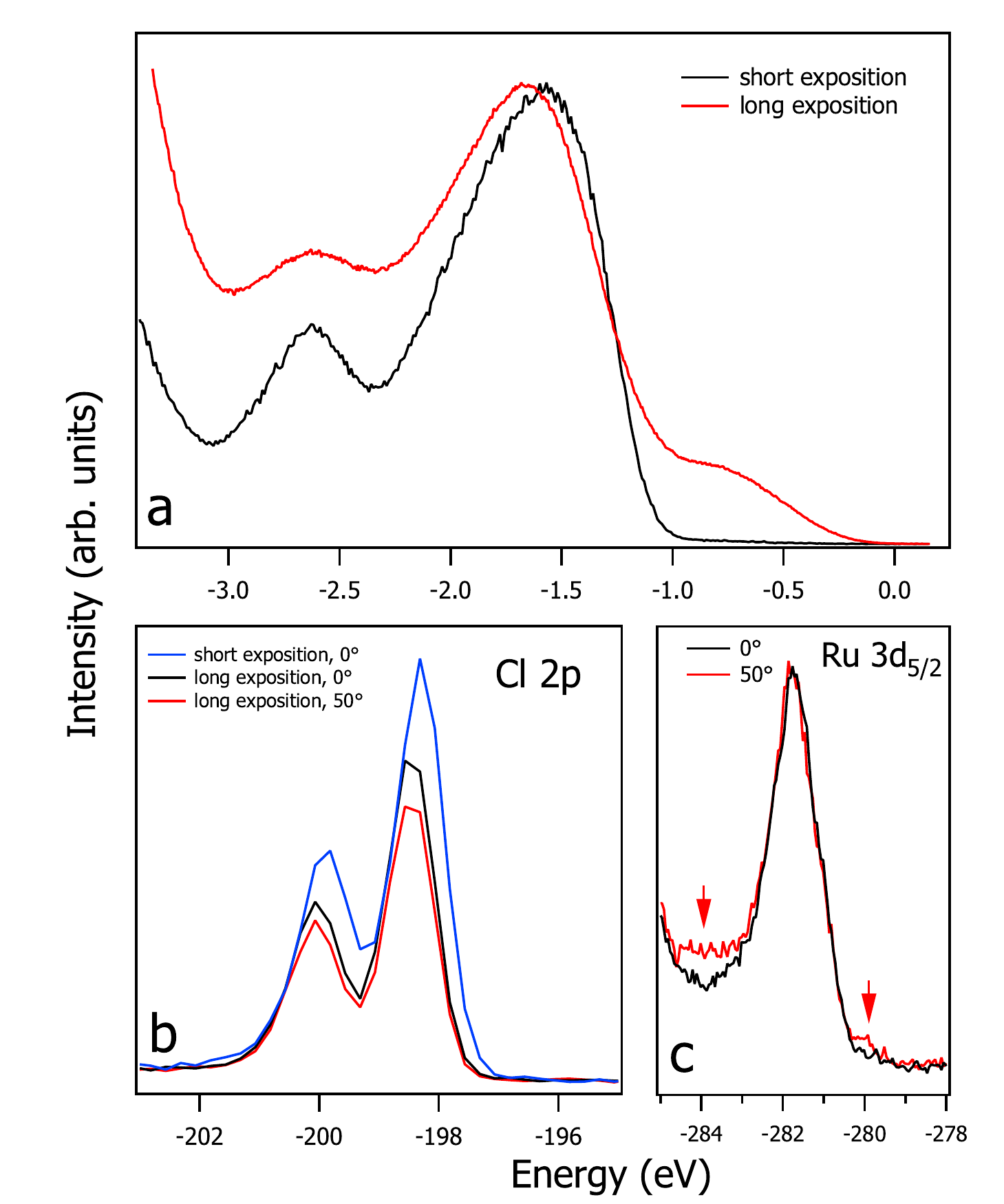}%
\caption{a) Valence band measured with h$\nu$=21.2 eV photon energy after short and long exposition. b) Cl 2p core level. c) Ru 3d$_{3/2}$.\label{}}%
\end{figure}

Core level spectroscopy is also a means for controlling the sample stoichiometry. Fig. 6b shows the Cl 2\textit{p} line for varying exposure times and emission angle normalized to the Ru 3$p$ line at higher energy (not shown). The intensity decreases for long exposure and when the emission angle increases, i.e. when the measurement becomes more surface sensitive. Additionally the peaks shift slightly to higher energies with exposure time. 
Obviously, the surface Cl content is not stable under illumination and suffers from desorption. This has severe consequences for the low energy electronic structure as shown in Fig. 6a for a surface sensitive valence band spectrum measured with h$\nu$ = 21.2 eV. A low energy shoulder develops and shifts \textit{E}$_{VBO}$ substantially towards \textit{E}$_F$ and reduces thereby the gap. The feature is likely of Ru 4\textit{d} character because it has an identical dependence on photon energy as the Ru 4\textit{d} peak. Moreover, a slight up-shift of the exposed spectra is seen also in the valence band. The features become broader and lose dispersion. The lineshape of the XPS Ru peaks is relatively stable, only small changes appear for the most surface sensitive measurements (see Fig. 6c). Also the XPS valence band hardly changes.
Cl photoreduction has been reported before for RuCl$_3$ \cite{Morgan2015}.
The Cl desorption will cause disorder at the surface which accounts for line broadening and loss of dispersion. It may also reduce Ru$^{3+}$ to lower oxidation states, which tends to shift spectral features to lower energy in agreement with the low energy shoulder developing in the valence band. The down-shift of the core-levels, on the other hand, could be a band filling effect.
The limited surface stability of \rc\ has to be taken into account for the interpretation of photoemission experiments and, possibly, surface sensitive techniques in general.   

In summary, we have investigated the electronic structure of \rc\ in detail. The angle-integrated valence band PES shows a main Ru 4\textit{d} contribution at \textit{E} = -1.6 eV with a width of \textit{W}$_{4\textit{d}}$ = 0.75 eV. The Fermi level must be pinned close to the bottom of the conduction band in our samples. The Cl 3\textit{p} derived states are located at higher energies and have a much larger width. In the angle-resolved mode, it becomes clear that the Cl 3\textit{p} states show also a much larger dispersion than the Ru 4\textit{d} bands. The latter disperse on the order of 200 meV only. All this is well described by DFT calculations using \textit{U}$_{eff}$ = 1.6 eV exhibiting a clear correspondence to the generic \textit{J}$_{eff}$ description of local cubic systems with large spin-orbit coupling. 
From EELS measurements the direct gap is \textit{E}$_g$ = 1.2 eV with a very sharp gap excitation mode. The crystal field splitting 10\textit{Dq} is about 2.2 eV and the charge transfer excitations span a broad energy range in the loss function reflecting their large width in the PES valence band. The gap value and the width of LHB and UHB are consistent with \textit{U}$_{eff}$ = 1.6 eV. 
The validity of the electronic parameters \textit{U}$_{eff}$, 10\textit{Dq} and $\Delta$ is confirmed by core level spectroscopy in combination with multiplet-calculations. The splitting of the 3\textit{p}$_{3/2}$ line into two components seen in EELS but not in XPS indicates the relevance of the SOC for the 4\textit{d}-electrons. 
Finally, XPS clearly shows a light induced loss of Cl at the surface, which alters the stoichiometry with time and substantially influences the valence band shape.

The above observations convey the picture of a Mott insulator whose low-energy structure is dictated by a mixture of the local cubic symmetry and spin-orbit coupling which might give rise to exotic magnetic groundstates.

\bibliography{D:/Literature/RuCl3}

\end{document}